\documentclass[aps,twocolumn,amsmath,amssymb,floatfix,pra,reprint,footinbib,superscriptaddress]{revtex4-1}

\usepackage{graphicx}
\usepackage{epstopdf}

\usepackage[applemac]{inputenc}
\usepackage[T1]{fontenc}
\usepackage{lmodern}
\usepackage[english]{babel}
\usepackage{ae}
\usepackage{units}
\usepackage{color}
\usepackage{url}

\usepackage{amsmath,amssymb,natbib,bm}
\usepackage{psfrag}
\usepackage{amssymb}
\usepackage{amsthm}
\usepackage{mathrsfs}

\usepackage{fixltx2e}
\usepackage{fixmath}
\usepackage{booktabs}

\usepackage{slashed}

\usepackage[americaninductors]{circuitikz}
\usepackage{tikz}
\usetikzlibrary{arrows}

\usepackage{dblfloatfix}

\usepackage[colorlinks]{hyperref}
\hypersetup{%
	plainpages=true,
	breaklinks=true,
	hypertexnames=false,
	pageanchor=true,
	colorlinks=true,
	linkcolor={blue},
	citecolor={red},
	urlcolor={blue},
	anchorcolor={black}
}

\newcommand{\be}{\begin{equation}}
\newcommand{\ee}{\end{equation}}
\newcommand{\bea}{\begin{eqnarray}}
\newcommand{\eea}{\end{eqnarray}}

\newcommand{\figref}[1]{\mbox{Fig.~\ref{#1}}}

\newcommand{\secref}[1]{\mbox{Section~\ref{#1}}}

\renewcommand{\eqref}[1]{\mbox{Eq.~(\ref{#1})}}

\begin{document}
\title{Topological edge states and pumping in a chain of coupled superconducting qubits}
\author{Xiu Gu}
\affiliation{Institute of Microelectronics, Tsinghua University, Beijing 100084, China}
\author{Shu Chen}
\affiliation{Beijing National Laboratory for Condensed Matter Physics, Institute of Physics, Chinese Academy of Sciences, Beijing 100190, China}
\author{Yu-xi Liu}
\email{yuxiliu@mail.tsinghua.edu.cn}
\affiliation{Institute of Microelectronics, Tsinghua University, Beijing 100084, China}
\affiliation{Tsinghua National Laboratory for Information Science and Technology
	(TNList), Beijing 100084, China}
\date{\today}

\begin{abstract}
Topological insulators have inspired the study with various quantum simulators. Exploiting the tunability of the qubit frequency and qubit-qubit coupling, we show that a superconducting qubit chain can simulate various topological band models. When the system is restricted to the single-spin excitation subspace, the Su-Schrieffer-Heeger (SSH) model can be equivalently simulated by alternating the coupling strength between neighboring qubits. The existence of topological edge states in this qubit chain is demonstrated in the quench dynamics after the first qubit is excited. This excitation propagates along the chain where the qubit-qubit coupling is homogeneous. In contrast, in our qubit chain, the spin-up state localizes at the first qubit and the rest qubits remain in the spin-down state. We further show that the spin-up state can be transported along the chain by modulating the coupling strengths and the qubit frequencies.
This demonstrates adiabatic pumping based on the Rice-Mele model. Moreover, we also discuss possible ways to construct other topological models with different topological phenomena within the current technology of superconducting qubits.
\end{abstract}

\maketitle

\section{Introduction}

Topology is no longer a pure mathematical discipline. It has been linked to many areas of physics. A revolution of topological physics has taken place in topological matter. In particular, the discovery of topological insulators~\cite{Hasan2010} triggered a wider study in topological phases of matter. Topological insulators, topological superconductors, and topological semimetals are just a few examples. It is known that the topological power stems from its global geometric properties characterized by topological invariant numbers. With these topological protections come many possible applications, for example, topology was introduced to solve the decoherence problem of quantum computation. In topological quantum computation~\cite{Stern2013}, the non-Abelian states of matter are used to encode and manipulate quantum information in a nonlocal manner. These non-local global states are topologically protected, and are more robust against the decoherence of qubit states or local impurities of quantum computational devices.

We know that many of the topological phenomena were demonstrated in crystals or other condensed matter systems by virtue of certain symmetries of these systems. However, topological physics is not only limited to condensed matter systems, but also has been applied to photonic systems, ultracold atoms, and ultracold gases in optical lattices~\cite{Lu2014a,Chien2015,Goldman2016}. These easily controlled or tuned systems enhance the possibility to create and probe new topological phases. Furthermore, inspired by quantum computing, in which the qubits and their couplings can be controlled or tuned, topological physics is studied via quantum computational devices. The reason is twofold. First, some exotic topological states, which are not easy to find in natural systems, may be created and probed by artificially designing and fabricating on-demand quantum computational devices. Second, some topological states, which are experimentally difficult to create in natural systems, may be simulated via quantum simulators~\cite{Georgescu2014}.

The simplest model exhibiting topological characters is the Su-Schrieffer-Heeger (SSH) model~\cite{Su1979,Su1980,Heeger1988,Asboth2016}. It has been extensively studied by theorists~\cite{Takayama1980,Jackiw1976,Ruostekoski2002,Li2014topo} and attracted different experimental platforms (e.g., in Refs.~\cite{Poli2014,Kitagawa2011,Cardano2017,Atala2013,Meier2016,Lohse2015,Nakajima2016}). For example, in cold atoms, the topological invariant of the one dimensional band, also known as the Zak phase~\cite{Zak1989},  was measured~\cite{Atala2013}. Because of the bulk-edge correspondence, the band invariant is associated with the existence of edge states. The edge signal is not easy to resolve from the bulk in the real space lattice of cold atoms. Recently in the momentum space of cold atoms, the dynamics of edge states was probed~\cite{Meier2016}. The quantized transport of particles, known as the Thouless pump~\cite{Thouless1983}, was also demonstrated in cold atoms by modulating the on-site potential and coupling strength of the SSH model~\cite{Lohse2015,Nakajima2016}.

Recently, topological physics is explored through superconducting quantum circuits (or superconducting artificial atoms)~\cite{You2011,Devoret2013,Gu2017}. Unlike natural atoms, these circuits can be fabricated with well-tailored characteristic frequencies and other parameters. The exquisite control of superconducting quantum circuits makes it possible to simulate topological band models on a single superconducting qubit.  This is achieved by mapping the bulk momentum space of a topological band model,  onto the parameter space of a spin in an external magnetic field~\cite{Gritsev2012}.  The Berry phase was first measured in a single superconducting qubit~\cite{Leek2007,Berger2012,Berger2013,Schroer2014,Zhang2017}.  Via Berry phase, topological invariants characterizing the band properties were also measured~\cite{Roushan2014,Flurin2016,Ramasesh2017}.  The space-time inversion symmetric topological semimetal~\cite{Tan2017} and topological Maxwell metal bands~\cite{Tan2017b} were also simulated in a single
superconducting qubit circuit. Experimental efforts are now directed to large scale of superconducting qubits. As an initial step towards realizing fractional
quantum Hall effect, anyonic fractional statistical behavior is emulated in supercondcuting circuits with four qubits coupled via
a quantized microwave field~\cite{Zhong2016}. Also, directional transport of photons was observed on a unit cell formed by three superconducting qubits~\cite{Roushan2016}.
In this design, qubits play the role of the lattice sites, whereas the synthetic topological materials are made of photons. There are various interesting theoretical proposals to
study topological photonic systems based on circuit QED structures~\cite{Koch2010,Nunnenkamp2011,Mei2015,Mei2016,Yang2016a,Tangpanitanon2016} . 

Here, rather than using microwave photons coupled by superconducting qubits, we propose to simulate topological physics with a chain of coupled superconducting qubits.  As a simulator of spin physics, the coupled superconducting qubits are
widely studied~\cite{Levitov2001,Tian2010,Johnson2011a,Viehmann2013,Viehmann2013a}.
For instance, quantum annealing was demonstrated experimentally on an Ising spin chain comprised of eight superconducting flux qubits~\cite{Johnson2011a}. Due to the improved controllability and fabrication technique of superconducting circuits, it becomes accessible to fabricate tens of qubits with various types of couplings. The qubit frequency and qubit-qubit coupling strengths can all be tuned in situ, making the whole superconducting qubit chain versatile enough to simulate topological models~\cite{Asboth2016,Harper1955,Aubry1980}.
Inspired by the studies in other systems (e.g., In Refs.~\cite{Poli2014,Kitagawa2011,Cardano2017,Atala2013,Meier2016,Lohse2015,Nakajima2016}),
we here study topological edges states and pumping by constructing the SSH model~\cite{Heeger1988,Asboth2016} using, e.g., gap-tunable flux qubit circuits~\cite{Paauw2009,Paauw2009Thesis,Schwarz2013}. 

The paper is organized as follows. In~\secref{sec:model}, we briefly introduce the gap-tunable flux qubit circuit. In~\secref{sec:chain},  we present a theoretical design for superconducting qubit chain, with either fixed or time modulated coupling, composed of gap-tunable flux qubits. We then show that this spin chain model can be mapped to SSH model or Rice-Mele model when restricted to single excitation subspace. In~\secref{sec:dynamics}, the quench dynamics of the SSH chain and the transverse Ising chain are compared. As we can see, the existence of edge states reveals itself as a soliton localized at the very end of the chain. In~\secref{sec:pumping}.
we show that the soliton can be transported from the first qubit to the last one by adiabatic pumping.  In~\secref{sec:summary}, we summary our results and further discuss possible demonstration on topological physics using superconducting qubit circuits. At the same time, we also show that the Hamiltonian constructed by the controllable superconducting qubit circuits can also be used to discuss other model of the condensed matter physics. For the completeness of the paper, we also give a detailed superconducting circuit analysis for the spin chain in the appendix.

\section{Gap-tunable flux qubit}\label{sec:model}

\begin{figure}[hbt]
\includegraphics[width=6cm]{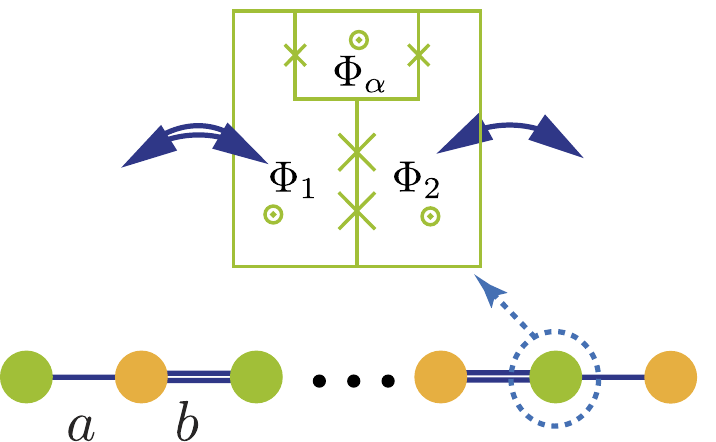}
\caption{Schematic diagram for a coupled qubit chain (lower panel) constructed by gap-tunable flux qubits with a gradiometric design (top panel).
 The $2L$ identical qubits are denoted by circles with green (orange) for odd (even) sites.
The double-line connecting the qubits denotes the coupling strength $b$, however single-line denotes the coupling strength $a$. The magnetic
flux threading the $\alpha$-loop is denoted by $\Phi_{\alpha}$.
 $\Phi_{1}$ and $\Phi_{2}$ denote magnetic fluxes through the left and right of the main loop.
 The signs $\bigodot$ denote that the magnetic fluxes are directed outside.
}\label{fig:system}
\end{figure}

As schematically shown in the lower panel of~\figref{fig:system}, we study a superconducting quantum circuit, in which $2L$ identical superconducting qubits are coupled to form a chain with alternating coupling strengths. That is, the coupling strength between the qubit (marked in green) on the odd sites and its right neighbor (marked in orange) is $a$, while the coupling strength between the qubit (marked in orange) on the even sites is coupled to its right neighbor (marked in green) with an amplitude of $b$. In principle, such qubit chain can be constructed by any type of the superconducting qubit circuit, e.g., flux qubits~\cite{Orlando1999,Liu2005a}, transmon~\cite{Koch2007},  xmon~\cite{Barends2013} and gmon~\cite{Chen2014,Geller2014}. But for concreteness of the discussions, we assume that each superconducting qubit in the chain is implemented by a gap-tunable  flux qubit~\cite{Paauw2009,Fedorov2010,Schwarz2013}, as schematically shown in the top panel of Fig.~\ref{fig:system} and is further explained in Fig.~\ref{fig:qubit} in the Appendix. For the completeness of the paper, below, we just briefly introduce the gap-tunable flux qubit circuit.

A gap-tunable flux qubit~\cite{Paauw2009,Paauw2009Thesis,Schwarz2013} is a variation of a three-junction flux qubit~\cite{Orlando1999,Liu2005a}. It replaces the smaller junction in the three-junction flux qubit with a superconducting quantum interference device (SQUID), which is equivalent to a single junction. The SQUID loop is referred to as $\alpha$-loop. An externally controllable flux $\Phi_{\alpha}$ applied to the $\alpha$-loop can change the Josephson energy of the smaller junction, and the ratio between the larger junctions and the smaller junction. This directly results in tunable tunneling between two potential wells of the flux qubit~\cite{Orlando1999}.  Also to keep $\Phi_{\alpha}$ from affecting the biased flux threading the main loop, a gradiometric design is adopted. In this $8$-shaped design, the central current of the three-junction flux qubit is split into two opposite running currents through two small loops. The magnetic flux generated by these two currents cancel each other in the main loop, thus independent control over both fluxes in $\alpha$-loop and main loop is ensured. Magnetic fluxes $\Phi_{1}$ and $\Phi_{2}$, applied to two small loops of the qubit, are used to tune the potential well energy of the qubit. Because both the tunneling and the potential well energies can be tuned in the gap-tunable flux qubit, we have a fully controllable Hamiltonian. Below, we use reduced magnetic fluxes $f_{\alpha}=\Phi_{\alpha}/\Phi_{0}$, $f_{\epsilon_{1}}=\Phi_{1}/\Phi_{0}$, and $f_{\epsilon_{2}}=\Phi_{2}/\Phi_{0}$. Here $\Phi_0$ is the flux quanta.

We define the flux difference between the two loop halves of the gradiometer as $f_\epsilon=f_{\epsilon_{1}}-f_{\epsilon_{2}}$. At the optimal point where $f_\epsilon=0$,  the low-frequency effect of the environmental magnetic flux reaches minimum, and the two lowest-energy states of the flux qubit are the supercurrents  states $\pm I_{p}$ circulating in opposite directions in the
$f_{\epsilon}$ loop. In the persistent current states basis,  e.g., the anticlockwise current state $|\circlearrowleft\rangle$ and clockwise current state $|\circlearrowright\rangle$, the Hamiltonian of the qubit takes the form
\begin{equation}\label{eq:1}
H=-\frac{1}{2}(\epsilon\,\sigma_z+\delta\,\sigma_x),
\end{equation}
where $\sigma_z$ and $\sigma_x$ are Pauli matrices, the parameter $\delta$ is the tunneling energy between the states of two potential wells and can be tuned by the
reduced magnetic flux $f_{\alpha}$. The parameter $\epsilon(f_\epsilon,f_\alpha)=2f_\epsilon\Phi_0I_{p}(f_\alpha)$ is the energy bias, which can be tuned by both
$f_{\alpha}$ and $f_\epsilon$.

In the qubit basis $|g\rangle$ and $|e\rangle$ obtained by diagonalizing the Hamiltonian in Eq.~(\ref{eq:1}), the qubit eigenstates have eigenenergies $\pm \omega/2$ with transition frequency $\omega=\sqrt{\epsilon^2+\delta^2}$ between two eigenstates. Hereafter we take $\hbar=1$ and the qubit Hamiltonian in Eq.~(\ref{eq:1}) is usually written as $H=\omega\sigma_{z}/2$, here $\sigma_{z}$ is redefined in the diagonalized basis $|g\rangle$ and $|e\rangle$. It is obvious that $\omega$ can be tuned by both $f_{\alpha}$ and $f_\epsilon$. Moreover, in contrast to the three-junction qubit, in which the transition frequency $\omega$ is  fixed at the optimal point, the gap-tunable flux qubit allows for independent control over the qubit frequency without affecting the bias point, that is, $\omega$ can be tuned by $f_{\alpha}$ even at the optimal point with $f_\epsilon=0$. This is demonstrated in detail in the appendix. Thus the gap-tunable flux qubits provide us long coherence time and easy control at the optimal point.

\section{Superconducting qubit chains and SSH model}\label{sec:chain}

As shown in Fig.~\ref{fig:system}, there are four current loops in the gap-tunable flux qubit, thus different types of tunable couplings can be created via different loops. For example, the longitudinal coupling $\sigma_j^z\sigma_{j+1}^z$ between the $j$th and $(j+1)$th qubits in the qubit chain can be realized by inductively coupling them through their $\alpha$-loops~\cite{Billangeon2015}. However, in this paper, we mainly focus on the transverse coupling $\sigma_j^x\sigma_{j+1}^x$ between the qubits via $\epsilon$-loops. Hereafter, $\sigma_{j}^{x}$, $\sigma_{j}^{y}$, and $\sigma_{j}^{z}$ are used to denote Pauli operators of the $j$th qubit with the ladder operators $\sigma_j^\pm=(\sigma_j^x\pm i\sigma_j^y)/2$, which are defined in the qubit basis $|g\rangle_{j}$ and $|e\rangle_{j}$ of the $j$th qubit.

To demonstrate different topological physics with controllable superconducting flux qubit circuits, in this section, we mainly show how to realize two coupling mechanism between gap-tunable flux qubits. (i) The qubits are directly coupled to each other, with fixed coupling strengths between qubits, but the frequencies of qubits can be either modulated or fixed. (ii) The qubits are indirectly coupled to each other via a coupler, thus both the qubit frequencies and the coupling strengths between qubits can be modulated.

\subsection{Qubit chain with fixed coupling strength and tunable frequencies of qubits}

\begin{figure}[hbt]
\includegraphics[width=8cm]{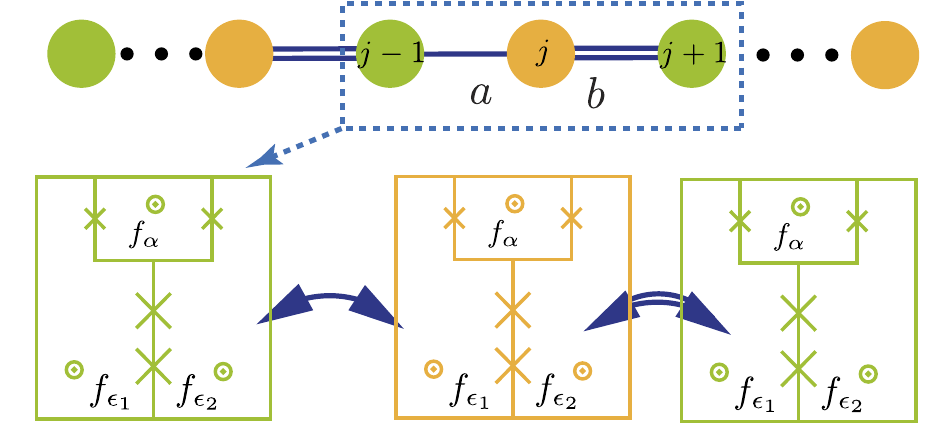}
\caption{Schematic diagram for the chain of coupled gap-tunable flux qubits with alternating coupling strengths $a$ and $b$ (Top panel).
As shown in the lower panel, each qubit is coupled to
its nearest-neighbors via the flux $f_{\epsilon_i}$. The different coupling strengths $a$ and $b$ are created by
varying the spacing between the qubits. The frequency of the qubit can be tuned in situ by the magnetic frustration
$f_\alpha$ threading the $\alpha$-loop.
}\label{fig:systemA}
\end{figure}

As schematically shown in ~\figref{fig:systemA}, we first study the case that the qubits are directly coupled to each other, that is, $2L$ identical gap-tunable flux qubits in the chain are directly coupled to each other through their mutual inductance via $\epsilon$-loops. We assume that each qubit is only coupled to its nearest neighbors and the couplings to other qubits are neglected, then the coupling strength between $j$th and $(j+1)$th qubits can be obtained by $J = MI_{pj}I_{p(j+1)}$. Here $M$ is the mutual inductance between  the $j$th and $(j+1)$th qubit loops, $I_{pj}$ and $I_{p(j+1)}$ are the current circulating main loops of the $j$th and $(j+1)$th qubits. Projecting $J$ onto the eigenstates of the $j$th and $(j+1)$th qubits, i.e., the states $|g\rangle_{j}$, $|g\rangle_{j+1}$, $|e\rangle_{j}$, and $|e\rangle_{j+1}$, we can obtain the coupled Hamiltonian, which includes the longitudinal coupling term $\sigma_j^z\sigma_{j+1}^z$ with the coefficient $g^{zz}_{j,j+1}$, transverse coupling term $\sigma_j^x\sigma_{j+1}^x$ with the coefficient $g^{xx}_{j,j+1}$ , and cross coupling terms, e.g., $\sigma_j^x\sigma_{j+1}^z$ with the coefficient $g^{xz}_{j,j+1}$.  Detailed analysis in the Appendix shows that the interaction via the main loops only results in transverse coupling when the qubits work at the optimal point, while the coefficients $g^{zz}_{j,j+1}$ and $g^{zx}_{j,j+1}$ of the longitudinal and cross couplings are zero.

 To make the qubit have long coherence, we assume that all gap-tunable qubits in the chain work at the optimal point in the following discussions, then there is only transverse coupling between the qubits.
 As shown in the Appendix, the coupling coefficient $g^{xx}_{j,j+1}$ of the transverse coupling between the $j$th and $(j+1)$th qubits is written as $g^{xx}_{j,j+1}=Mg^j_{\epsilon,\bot} g^{j+1}_{\epsilon,\bot}$, with $g^j_{\epsilon,_\bot}={_{j}\langle e|I_{pj}|g\rangle}_j$ and  $g^{(j+1)}_{\epsilon,_\bot}={_{j+1}\langle e|I_{p(j+1)}|g\rangle}_{j+1}$. To create alternating coupling pattern as schematically shown in~\figref{fig:system}, the spacings between the qubits need to be varied respectively in order to alter $M$. This is experimentally accessible with current technology of the superconducting qubit circuits. The qubit chain is fabricated such that
 $a\equiv g^{xx}_{(2m-1),2m}$ and  $b\equiv g^{xx}_{2m,(2m+1)}$ with $m=1,\,2,\cdots, L$. We note that $b=0$ when $m=L$. Then, the Hamiltonian of the coupled-qubit chain can be written as
\bea
\label{eq:SSH}
H&=&\sum_{j=1}^{2L}\frac{\omega}{2}(\sigma_{j}^{z}+1)+\sum\limits_{{l\in \mathrm{odd}}}^{2L}a(\sigma_{j}^{+}\sigma_{j+1}^{-}+{\text H.c.})\nonumber\\
 &+& \sum\limits_{{j\in \mathrm{even}}}^{2L}b(\sigma_{j}^{+}\sigma_{j+1}^{-}+{\text H.c.}),\nonumber\\
\eea
with $\omega$ the frequency  of the qubit. Because we assume that all qubits are identical, thus the qubits have the same frequency $\omega$. That is, the qubits resonantly interact with each other, and then we can make rotating wave approximation such that the transverse coupling term $\sigma_{j}^{x}\sigma_{j+1}^{x}$ between the $j$th and $(j+1)$th qubits is approximated as $\sigma_{j}^{x}\sigma_{j+1}^{x}\approx (\sigma_{l}^{+}\sigma_{l+1}^{-}+\sigma_{j}^{-}\sigma_{j+1}^{+})$.

As shown in Fig.~\ref{fig:systemA} and discussed in Section~\ref{sec:model}, when the qubit works at the optimal point, the frequency modulation can be only done by applying the time-dependent magnetic flux through the $\alpha$-loop. However, when all the qubits do not work at the optimal point, although the frequency of each qubit and the coupling strength between the qubits can be modulated by a time-dependent magnetic flux applied through the main loop of each qubit~\cite{Liu2006}, the coherence of the qubits is not satisfying.

\subsection{Qubit chain with both tunable coupling strengths and frequencies of qubits}

\begin{figure}[hbt]
\includegraphics[width=8cm]{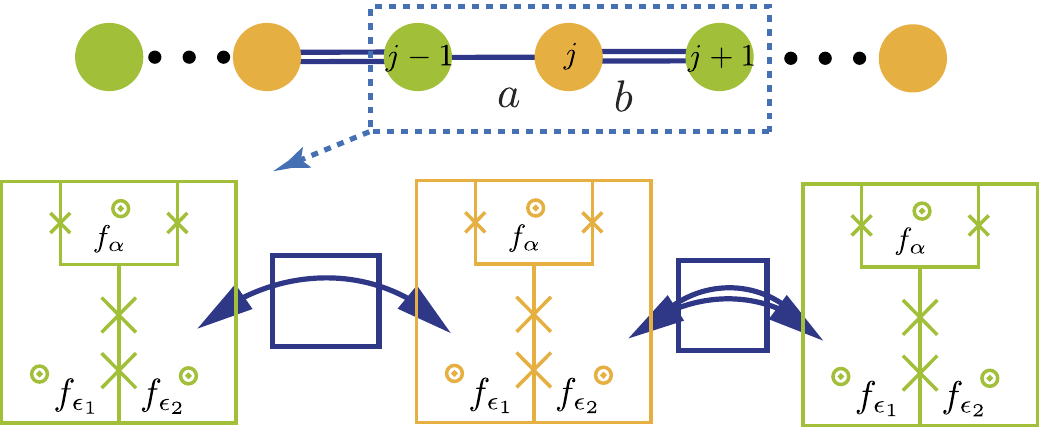}
\caption{Couplers (rectangle) are inserted between qubits in the chain, as shown in Fig.~2, to achieve time-dependent tunable coupling strengths. The couplers can be superconducting quantum interference devices, superconducting qubits, LC circuits, or other superconducting elements.
}\label{fig:systemB}
\end{figure}

To make sure the coupling strengths between qubits can be modulated and also the qubits work at the optimal point,  an additional coupler~\cite{Liu2006,Liu2007,Grajcar2006,Rigetti2005,Bertet2006,Niskanen2006}, as schematically shown in ~\figref{fig:systemB},
needs to be inserted between the qubits. This sort of coupler has been demonstrated in
the two-qubit case~\cite{Hime2006,Liu2006,Liu2007,Niskanen2007,Harris2007,VanDerPloeg2007,Allman2010,Bialczak2011}. In this case, using the same procedure for the directly coupled qubit chain in Fig.~\ref{fig:systemA}, we can write out the full Hamiltonian of the qubit chain coupled through the couplers. The modulated couplings of the Hamiltonian can be realized by individually applying the magnetic flux through the loop of each coupler. To obtain an effective Hamiltonian, we use two ways to eliminate the variables of the couplers. If the couplers work at the classical regime, and then the variables of the couplers can be adiabatically eliminated as in Ref.~\cite{Grajcar2006}. If the couplers work at the quantum regime, then we can eliminate the variables of the couplers by the large detuning approximation as in Refs.~\cite{Bertet2006,Niskanen2006}. In either way, the Hamiltonian in Eq.~(\ref{eq:SSH}) can be modified to
\bea
\label{eq:RM}
H&=&\sum\limits_{{j\in \mathrm{odd}}}^{2L}\left[\frac{\omega+{u}(t)}{2}(\sigma_{j}^{z}+1)+a(t)(\sigma_{j}^{+}\sigma_{j+1}^{-}+\sigma_{j}^{-}\sigma_{j+1}^{+})\right]\nonumber\\
&+&\sum\limits_{{j\in\mathrm{even}}}^{2L}\left[\frac{\omega-{u}(t)}{2}(\sigma_{j}^{z}+1)+b(t)(\sigma_{j}^{+}\sigma_{j+1}^{-}
+\sigma_{j}^{-}\sigma_{j+1}^{+})\right].\nonumber\\
\eea
Here, we emphasize that the on-site staggered potential $u(t)$ can be experimentally realized, as for the model shown in Fig.~\ref{fig:system} and discussed for Fig.~\ref{fig:systemA}, in superconducting flux qubit circuits by varying the magnetic flux through either the $\alpha$-loop or the main loop when the qubits do not work at the optimal point. However, if the qubits work at the optimal point discussed in this subsection, then the modulation $u(t)$ can only be realized by applying the magnetic flux through the $\alpha$-loop.

\section{Superconducting qubit chain and SSH model}

Let us now show how the Hamiltonians in Eq.~(\ref{eq:SSH}) can be mapped to the SSH model when restricted to the single-excitation subspace of the qubit chain. In most studies, the qubit operator is mapped onto the non-interacting fermions through
Jordan-Wigner transformation~\cite{Fradkin2013}.
However, here we find that the total spin excitation $\sum_{j=1}^{2L}(\sigma_{j}^{z}+1)$ commutes with Hamiltonian $H$  in both~\eqref{eq:SSH} and~\eqref{eq:RM}.  Thus the number of total excitations of the qubit chain is conserved.
In the following analysis, instead of resorting to the nonlocal Jordan-Wigner transformation, we restrict our study in the single-excitation subspace.
That is, only one qubit is excited in the $2L$ qubits. This can be done in superconducting quantum circuits due to the strong anharmonicy of the flux qubit. We define the basis
\begin{equation}\label{eq:3-1}
|e_{j}\rangle=|0,...,1_{j},0...\rangle,
\end{equation}
 where the $j$th superconducting qubit
is assumed in the spin-up state $|1\rangle$, while the others are assumed in the spin-down
states $|0\rangle$. Then in the subspace of only one excitation, the Hamiltonian $H$, e.g., in~\eqref{eq:SSH},  has the tridiagonal form
\be
H_{S}=\begin{bmatrix}
\omega & a &  & &&\\
a & \omega & b & &&\\
 & b & \omega & a&&\\
&  & \ddots & \ddots&\ddots&\\
&&&b&\omega& a\\
&&  &  & a & \omega\label{eq:matrix}
\end{bmatrix}.
\ee
Here, the subscript $S$ denotes the single-excitation.
After shifting the zero-energy point to $\omega$, the Hamiltonian in \eqref{eq:matrix} is equivalent to the SSH model~\cite{Asboth2016}, in which the single-particle Hamiltonian is given by
\begin{equation}\label{eq:4}
H_{\mathrm{SSH}}=\sum\left(aA_j^\dagger B_j+bA_j^\dagger B_{j-1}+\mathrm{H.c.}\right),
\end{equation}
where $A_j^\dagger$ ($B_j^\dagger$) is the particle creation operator on the site A (B) in the $j$th cell.
The SSH model describes a chain of dimers, each hosting two sites A and B.
The hopping strength within the unit cell is $a$, the intercell hopping amplitude is $b$. In our qubit chain as
schematically shown in~\figref{fig:system}, the odd (even) number of the qubits in Eq.~(\ref{eq:SSH}) corresponds to
the A (B) particles in Eq.~(\ref{eq:4}). Similarly, in the case of single-excitation, the Hamiltonian in Eq.~(\ref{eq:RM}) can also be mapped to one of Rice-Mele model to realize the topological pumping, which will be studied in Section~\ref{sec:pumping}.

\section{Topologically protected solitons}\label{sec:dynamics}

We now study how the proposed superconducting qubit chain with Hamiltonian in Eq.~(\ref{eq:SSH}) can be used to demonstrate topologically protected solitons when there is only a single-excitation in $2L$ superconducting qubits.  To clearly illustrate this, let us fix the coupling strength between the even qubit and its right neighbor, e.g., hereafter we take the coupling strength $b$ as a unit, i.e., $b=1$. Then we study how the hopping amplitude $a$,  the hopping strength between the odd qubit and its right neighbor, affects the energy spectrum and the topological properties of the chain.

\begin{figure}[h]
\includegraphics[width=\linewidth]{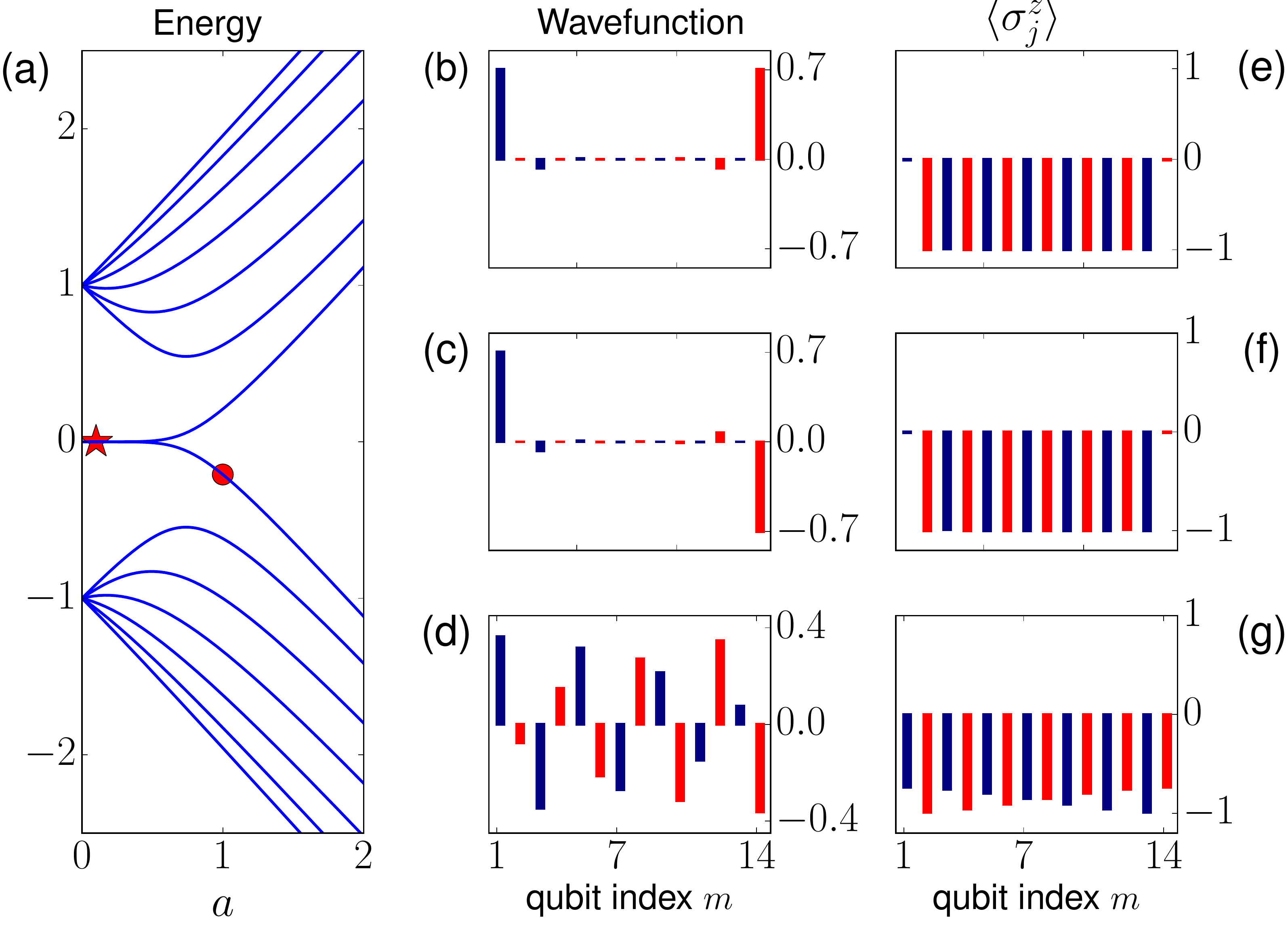}
\caption{Spectrum and wave functions of the Hamiltonian~(\ref{eq:matrix}). The total number of
qubits is assume as $14$. (a) Energy spectrum verses the coupling strength $a$ with $b=1$. Degenerate wave functions, corresponding
to the zero-energy mode with $a=0.1$ indicated by a star in (a), are shown in (b) and (c), respectively. A typical bulk state at
the point $a=1$, marked by a circle in (a), is plotted in (d). The observables $\langle\sigma_n^z\rangle$ of
the corresponding wave functions in (b), (c) and (d), are shown in (e), (f) and (g). All the navy blocks denote the amplitude on the odd
qubits, while the red blocks represent the even sites.}\label{fig:energylevel}
\end{figure}

In our simulation, for concreteness, the total number of superconducting qubits is set as $2L=14$. We plot, in~\figref{fig:energylevel}(a), the energy spectrum of the Hamiltonian in~\eqref{eq:matrix} and corresponding eigenfunctions $|\psi_l\rangle$ where $l=1,\cdots, 2L$ in the basis $|e_j\rangle$ shown in Eq.~(\ref{eq:3-1}).  They are plotted in Figs.~\ref{fig:energylevel}(b) and (c) with $a=0.1$, in Fig.~\ref{fig:energylevel}(d) with $a=1$.

Figure~\ref{fig:energylevel} bears several interesting features.
(i) Due to the bipartite lattice structure, the spectrum exhibits two band.
(ii) The spectrum is symmetric around zero. For any state with energy $E$, there is a partner with energy $-E$. This stems from the chiral symmetry of the SSH model~\cite{Asboth2016}.
(iii) For the $E\neq0$ states, \textit{all} the wave functions have support on both even and odd qubits, also known as the bulk states.  Figure~\ref{fig:energylevel}(d) shows a typical bulk state wave function corresponding to the point $a=b=1$ marked by a red point in~\figref{fig:energylevel} (a). (iiii) There is a zero energy ($E=0$) mode lying in the middle of the bulk gap, where $a<b$.
The zero-energy mode has two degenerate states. They are presented in~\figref{fig:energylevel} (b) and (c) corresponding to $a=0.1$ point marked by a star in~\figref{fig:energylevel} (a). The eigenfunctions are localized at the left and right edge, and decay exponentially towards the bulk.

The appearance of $E=0$ mode with localized eigenfunctions is the key feature of the topological phase when $a<b$.
The localized eigenfunctions shown in Figs.\ref{fig:energylevel} (b) and (c) are the superpositions $|L\rangle\pm|R\rangle$ of the left and right edge states $|L\rangle$ and $|R\rangle$. Here the left edge state is defined as
\begin{equation}
|L\rangle=\sum\limits_{j\in \mathrm{odd}} a_j|e_j\rangle,
\end{equation}
here $j$ is an odd number and $a_j$ is the amplitude on the odd qubits. Similarly, the right edge state is written as
\begin{equation}
|R\rangle=\sum\limits_{j\in \mathrm{even}} b_j|e_j\rangle,
\end{equation}
here $j$ is an even number and $b_m$ is the amplitude on the even qubits.
Note the vanishing amplitudes on the even (odd) site of the left (right) edge state are the consequence of the chiral symmetry.  The decay depth into the bulk is characterized by~\cite{Asboth2016}
\begin{equation}
|a_j|=|a_j|{\rm exp}\left(-\frac{j-1}{\xi}\right),
\end{equation}
where the localization length $\xi=(\ln |a|-\ln|b|)^{-1}$. When the ratio $b/a$ becomes appreciably large, the wave function will almost be confined at the first and last qubit.

The particle distributions in the eigenfunctions  $|\psi_l\rangle$ of the qubit chain can be measured via the variable $\sigma_{n}^z$ of each qubit, with the measurement result $\langle \psi_l|\sigma_n^z|\psi_l\rangle$. Here, the subscript $n$ denotes the $n$th qubit.  We note that the in the single-excitation subspace, the Pauli operator $\sigma_{n}^z$ can be expressed as $\sigma_n^z=2|e_n\rangle\langle e_n|-I$, where $I$ is the identity matrix. Although the qubits are coupled to each other, the excitation stays at the end of qubit chain as shown in~\figref{fig:energylevel} (e) and (f).

To demonstrate the existence of edge states, we study the quench dynamics after the first qubit is flipped.
That is, let us assume that all the qubits are initially in the spin-down state, then a $\pi$ pulse is applied to flip the first qubit. The dynamics of the qubit chain can be measured by
\be
\langle\sigma_n^z(t)\rangle=\langle e_1|e^{iH_\mathrm{S}t}\sigma_n^ze^{-iH_\mathrm{S}t}|e_1\rangle.
\ee
with the Hamiltonian given in Eq.~(\ref{eq:matrix}).
Below, we compare the dynamics of a topological SSH chain, described by the Hamiltonian in Eq.~(\ref{eq:matrix}) when $a=0.1$ and $b=1$, with a transverse spin chain, described by the Hamiltonian in Eq.~(\ref{eq:matrix}) when $a=b=1$. 

To model the small divergence of the qubits resulted from the sample fabrication, a random noise $\eta$  is introduced to the coupling strengths $a$ and $b$, as well as the frequency $\omega$ in Eq.~(\ref{eq:matrix}).  Here $\eta$  follows Gaussian distribution with mean value $0$, and a standard deviation $0.01$.  That is the fluctuations of the coupling strength and qubit frequencies are $10\%$ of the smaller coupling strength $a$.

\begin{figure}[hbt]
	\centering
	\includegraphics[width=8cm]{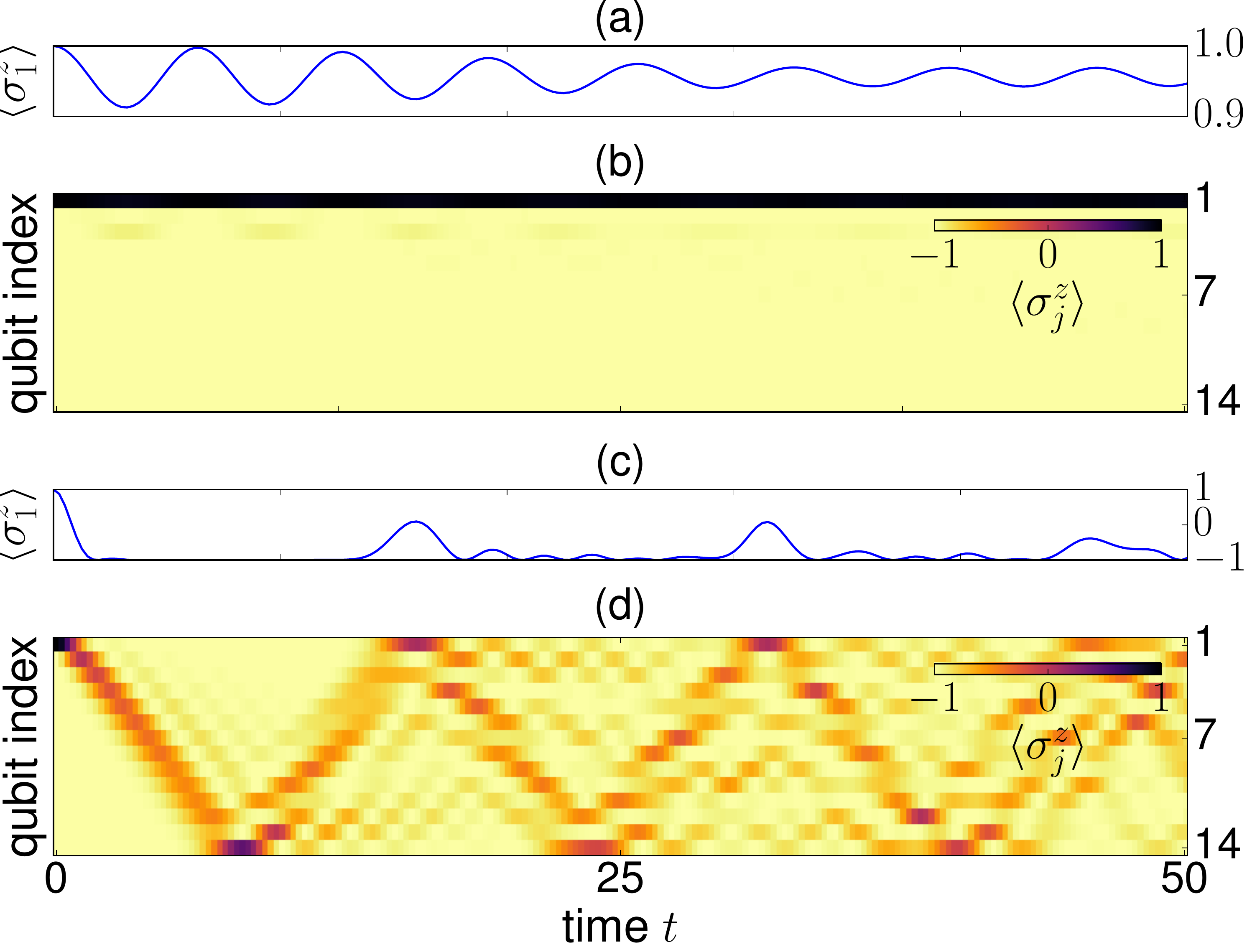}
	\caption{Time evolution of $\langle\sigma_{j}^z\rangle$, after the first qubit is flipped to the spin-up state. (a) and (c) show the time evolution $\langle\sigma_{1}^z\rangle$ of the first qubit.  (b) and (d) show the time evolution $\langle\sigma_{i}^z\rangle$ of all the qubits.  A random noise with an amplitude of $10\%$ of the coupling strength $a$ is added to the qubit frequency and coupling strength. The topological SSH chain with $a=0.1$ and $b=1$ is shown in (a) and (b). The transverse Ising chain with $a=b=1$ is plotted in (c) and (d). In our plot, the  qubit number in the chain is assumed as $14$.
	}\label{fig:trivial}
\end{figure}

For the topological SSH chain with $a=0.1$ and $b=1$ in the Hamiltonian of Eq.~(\ref{eq:matrix}), \figref{fig:trivial} (a) and ~\figref{fig:trivial} (b) show that the excitation remains as a soliton at the first qubit. This can be understood by the evolution of the wave function
\begin{equation}
|\psi(t)\rangle=\sum_ne^{-E_lt/\hbar}\langle \psi_l|e_1\rangle|\psi_l\rangle,
\end{equation}
where $|\psi_l\rangle$ denotes the $l$th eigenfunctions of the Hamiltonian in \eqref{eq:matrix} with corresponding eigenenergy $E_l$.  We start with the state $|e_1\rangle$ after the excitation is injected at the first qubit.  $|e_1\rangle$ has a substantial overlap with the degenerate edge states with corresponding eigenenergy $E=0$. This leads to a stationary state. Conversely, if we inject the excitation in the transverse Ising chain described by the Hamiltonian of Eq.~(\ref{eq:matrix}) with $a=b=1$, without localized edge states, it will quickly diffuse into the bulk. This can be seen from Figs.~\ref{fig:trivial}(c) and (d). The excitation at the first qubit quickly expands into the bulk, reaches the end of the qubit chain, and then is reflected back. The similar propagation of such excitations is demonstrated in Refs.~\cite{Viehmann2013,Viehmann2013a}.

In summary, due to the alternating coupling pattern $a<b$ of the whole chain, the soliton is topologically protected and robust against disorder. This is because the soliton resides in the gap, extra energy is required if we want to excite the soliton to other states. We have also shown the random noise added to the parameters in~\eqref{eq:SSH} have no appreciable influence on the soliton state.

Though the soliton appears at the very end of the chain, it can also be created at the interface between the topological phase with $a<b$ and topologically trivial phase with $a>b$. For example, as shown in Ref.~\cite{Estarellas2016}, if a defect is created at the center of a topological SSH chain, then a zero energy mode localizes at the defect.  Moreover, the defect can serve as a high-fidelity memory, which is topologically protected. An arbitrary state encoded with the presence and absence of the localized state can allow for perfect state transfer in the spin chain~\cite{Estarellas2016,Bose2003}.

\section{Pumping of an edge state}\label{sec:pumping}

We have seen in the previous section, once an excitation is injected at the edge, it will stay as a soliton. However, it is possible to transfer the soliton from the left to the right of the qubit chain by adiabatic pumping. The pumping can be realized if the staggered potential $u(t)$ and $-u(t)$ are added  respectively in the odd and even sites of the Hamiltonian in Eq.~(\ref{eq:SSH}), and also the coupling strengths $a$ and $b$ in the Hamiltonian of Eq.~(\ref{eq:SSH}) are changed to time-dependent $a(t)$ and $b(t)$. These kinds of the time modulations can be realized in superconducting qubit circuits with the Hamiltonian in Eq.~(\ref{eq:RM}) as discussed in Section~\ref{sec:chain}~B.

If we only consider the single-excitation, then as for discussion of the Hamiltonian in Eq.~(\ref{eq:matrix}) for the SSH model, the Hamiltonian in Eq.~(\ref{eq:RM}) can be reduced to the Hamiltonian of the Rice-Mele (RM) model~\cite{Rice1982,Asboth2016} with
\bea
H_{\mathrm{RM}}&=&\sum \left[a (t)A_n^\dagger B_n+b(t)A_n^\dagger B_{n-1}+\mathrm{H.c.}\right]\nonumber\\
&+& u (t)\sum(A_n^\dagger A_n -B_n^\dagger B_n),
\eea
where $A_n^\dagger$ ($B_n^\dagger$) is the particle creation operator on the site A (B) in the $n$th cell, $a(t)$ and $b(t)$ are the time dependent coupling strength, $u(t)$  is the staggered potential.
The degenerate point of the RM model in~\eqref{eq:RM} is given by $a=b$, $\mathrm{u}(t)\equiv 0$. This is the point where two bands touches, the topological and trivial phase of the SSH model are connected. The RM Hamiltonian can be continuously deformed along the time dependent pump sequence given by $u(t)$, $a(t)$ and $b(t)$. In superconducting quantum circuits, this can be done by varying the magnetic fluxes through the loops of the couplers and the $\alpha$ loop of the supercoducting qubits. As long as the time dependent path encircles the degenerate point with $a=b$ and $u=0$, all the Hamiltonians along the pathes are topologically equivalent~\cite{Asboth2016}. This gives us plenty of freedom to design the pump sequences. For example, we can demonstrate the topological pumping by simply choosing the coupling strengths $a(t)$ and $b(t)$ as well as the on-site potential $u(t)$ as
\bea\label{eq:pump}
a(t)&=&1-\cos\left(\frac{2\pi t}{T}\right),\nonumber\\
b(t)&=&1,\nonumber\\
\mathrm{u}(t)&=&-\omega\sin\left(\frac{2\pi t}{T}\right).
\eea
\begin{figure}[hbt]
	\centering
	\includegraphics[width=\linewidth]{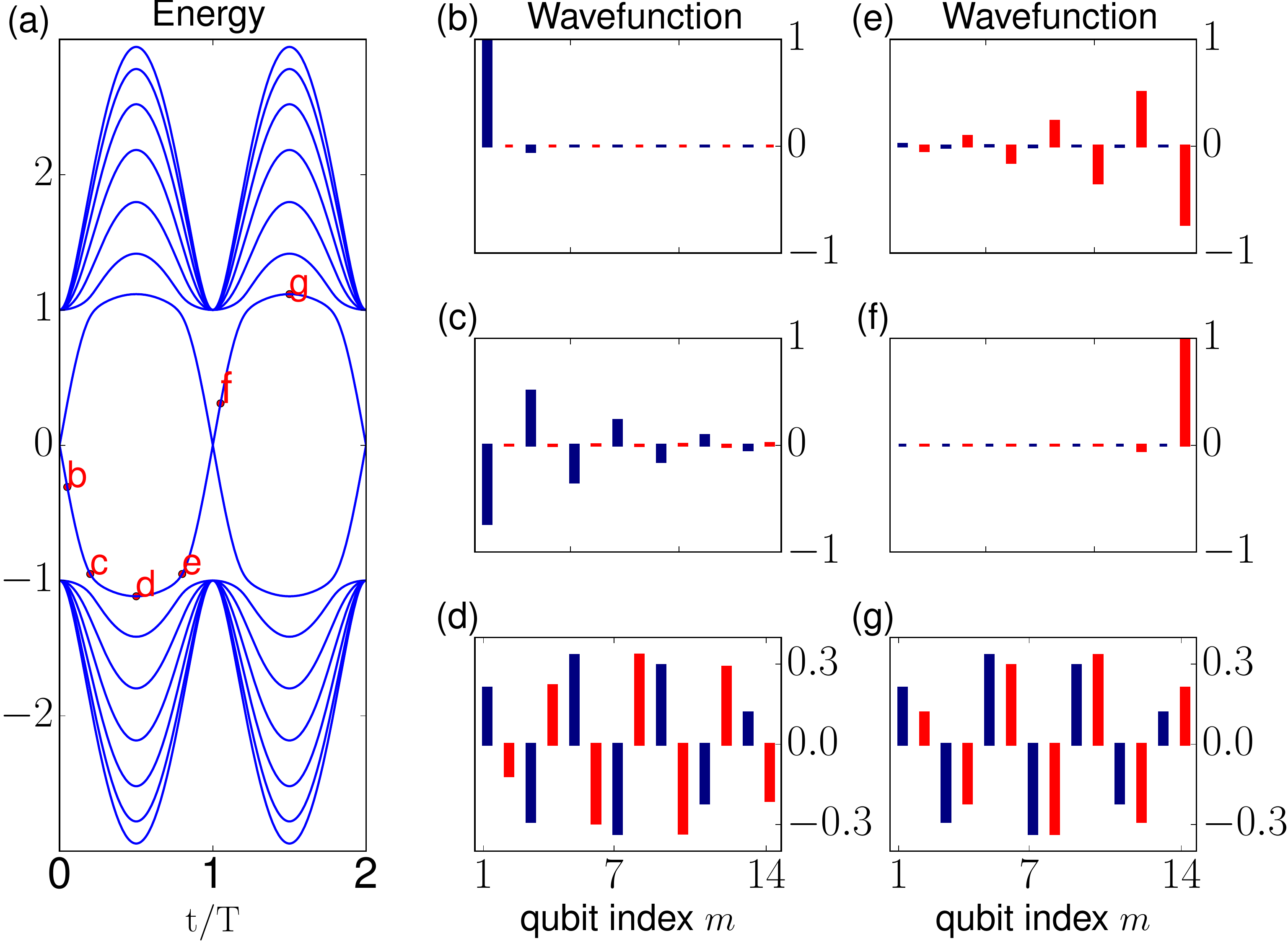}
	\caption{(a) Instantaneous spectrum of the Hamiltonian in Eq.~(\ref{eq:RM}) with the pumping sequence defined in~\eqref{eq:pump}.
	Corresponding wave functions are shown in (b) to (g). The chain consists of $2L=14$ qubits.
	}\label{fig:RM}
\end{figure}

The instantaneous spectrum of the Hamiltonian in~\eqref{eq:RM} is plotted as a function of $t$ in~\figref{fig:RM} (a). We can easily find that the system will stay in the same eigenstate as long as the adiabatic approximation holds. Through the wave functions at points b to f in~\figref{fig:RM}, we have shown how a left edge state is adiabatically pumped to the right during a pumping cycle.

The same with discussion of topologically protected solitons in~\secref{sec:dynamics}, a random noise $\eta$ is added to the coupling strength $a$, $b$ and the frequency $\omega$ in~\eqref{eq:RM},  where $\eta$ follows Gaussian distribution with mean value $0$, and a standard deviation $0.01b$.
To ensure the adiabatic limit, we set $T=100$. The chain is initialized in the all spin-down state. To prepare the left edge state, the first qubit is flipped by a $\pi$ pulse of the applied magnetic flux through the main loop of the qubit.

\begin{figure}[hbt]
	\centering
	\includegraphics[width=\linewidth]{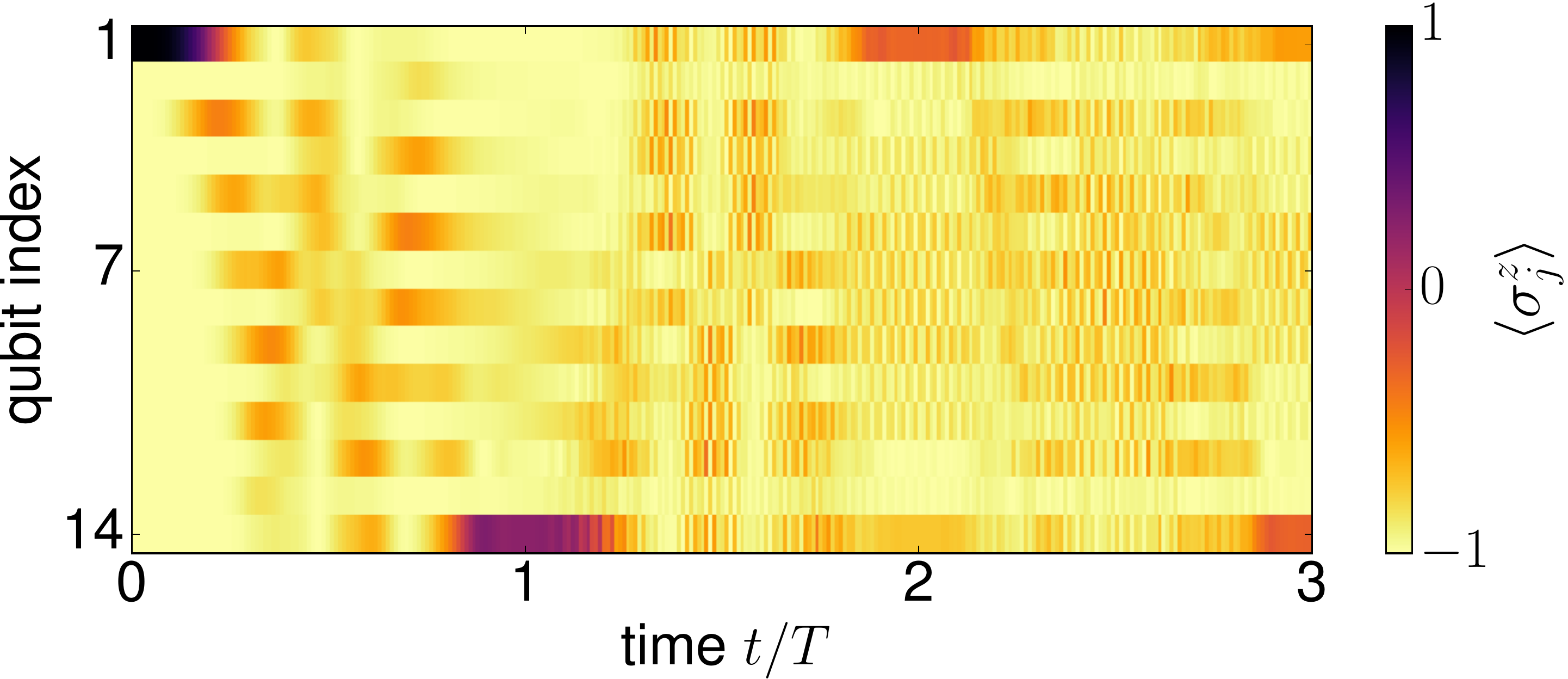}
	\caption{Time evolution of the qubit chain of $2L=14$ qubits. The time dependent pump sequence is defined in~\eqref{eq:pump}, with $T=100$. Random noise with an amplitude of $1\%$ of the coupling strength $b$ is added to the qubit frequency and qubit-qubit coupling.
	}\label{fig:pump}
\end{figure}

The dynamics of the time-dependent chain is solved numerically using Qutip~\cite{Qutip1,Qutip2}. For the first pumping cycle, the result shown in~\figref{fig:pump} is consistent with the adiabatic limit in~\figref{fig:RM}. The soliton first diffuses as the left edge state with vanishing amplitudes on the even sites. Then it is pushed into the bulk occupying both even and odd sites. After that, it reappears as the right edge state with vanishing amplitudes on the odd qubits. At the end of the first pumping cycle, the right edge state refocuses on the right end qubit. At $t=T$, Landau Zener transition occurs at the degenerate point, this results in the less well-resolved pattern in the following cycles.

In our pump result, only the edge mode is occupied initially, while the lower band is empty.
However, in the cold atom experiments~\cite{Lohse2015,Nakajima2016}, all the lower band is filled with the atoms, while the upper band is empty. During a pump cycle, each atom in the valence band is moved to the right by a single lattice constant. Or equivalently, the number of pumped particles through the cross section is one. This is determined by the Chern number of the associated band~\cite{Asboth2016}.
By promoting the periodic time $t$ to the wave-number, the adiabatic pump sequence in one dimension is equivalent to two-dimensional insulators.

\section{Further discussions and conclusions}\label{sec:summary}

We have proposed to simulate topologically protected solitons using a gap-tunable superconducting-qubit chain, in which the Hamiltonian is equivalent to the SSH model when the total excitations of the qubit chain is limited to one. We show that topological edge states can be directly probed through the quench dynamics of the chain after a spin-up state is injected. The spin-up injection at the localized edge state is robust against fluctuations, which can be used to store quantum information.
We further show an equivalence of the Rice-Mele model can be realized with the time modulated frequencies and the coupling strengths of the qubit chain, and the adiabatic pumping of an edge state can be realized in this time modulated qubit chain. In our numerical simulation, we take a larger number of the qubits in the chain, e.g., $14$ qubits, however, we find that the topological phenomena can also be demonstrated in such chain with even smaller qubit number, e.g., $8$ qubits. We also find that the localization can become more strong with the increase of the qubit number of the chain.

We note that our qubit chain can be used to implement pumping using one-dimensional Aubry-Andre-Harper (AAH) model~\cite{Harper1955,Aubry1980,Lang2012}, which is related to the well-known Hofstadter butterfly problem~\cite{Hofstadter1976} in two dimensions. In our proposal, the AAH model can be obtained through the Hamiltonian in Eq.~(\ref{eq:SSH}), in which the frequency $\omega$ of the $j$th qubit  needs to be modulated as $\omega \cos(2\pi j\alpha+t/T)$, where $\alpha$ is a rational (irrational) number. However, the qubit-qubit coupling strengths need to be changed to uniform, i.e., $a=b$. Experimentally, this can be realized by fabricating the chain of coupled superconducting flux qubits with uniform coupling strength, the frequency modulation can be done by applying the magnetic fields through the main loop of each qubit. However, if the coupling strengths of superconducting qubit chain are tunable, then we need only to tune all coupling strengths so that they equal to each other. We further note that the pumping of an edge state based on AAH model was realized in quasicrystals~\cite{Kraus2012}. Recently, the Hofstadter butterfly spectrum was observed in a chain of nine coupled gmon qubits~\cite{Roushan2017}.

We mention that the qubit chain can also be constructed using circuit QED system~\cite{Devoret2013,Gu2017}, where the qubit-qubit coupling can be mediated by the cavity fields. In this case, the cavity fields works as quantum couplers, thus the qubit-qubit coupling can be obtained by eliminating the cavity field with assumption that the qubits and the cavity fields are in the large detuning.

In summary, superconducting quantum circuits can artificially be designed according to the purpose of the experiments. In particular, the qubit frequencies and qubit-qubit couplings can be easily modulated or tuned. This opens many possibilities to simulate or demonstrate various topological physics of matter on demand with artificial designs. Moreover, the interaction between the single-mode cavity fields and topology matter might provide an opportunity to adjust the statistical or topological properties of the cavity fields via the topological properties of matter. These interesting physics are still under the study.

\begin{acknowledgments}
Y.X.L. acknowledges the support of the National Basic Research Program of
China Grant No. 2014CB921401 and the National Natural Science Foundation of
China under Grant No. 91321208. S.C. is supported by the National Key Research
and Development Program of China (2016YFA0300600), NSFC under Grants No. 11425419, No. 11374354 and No. 11174360.
\end{acknowledgments}

\appendix \label{appendix}

\section{Tunable qubit-qubit coupling}
In this section, the Hamiltonian of inductively coupled gap-tunable flux qubit chain (~\figref{fig:system}) will be derived. We will show $\omega$ and $a$, $b$ (in ~\eqref{eq:SSH}) can be tailored and tuned in situ by external fluxes.

\subsection{Gap-tunable flux qubit Hamiltonian}

\begin{figure}[hbt]
\includegraphics[width=8cm]{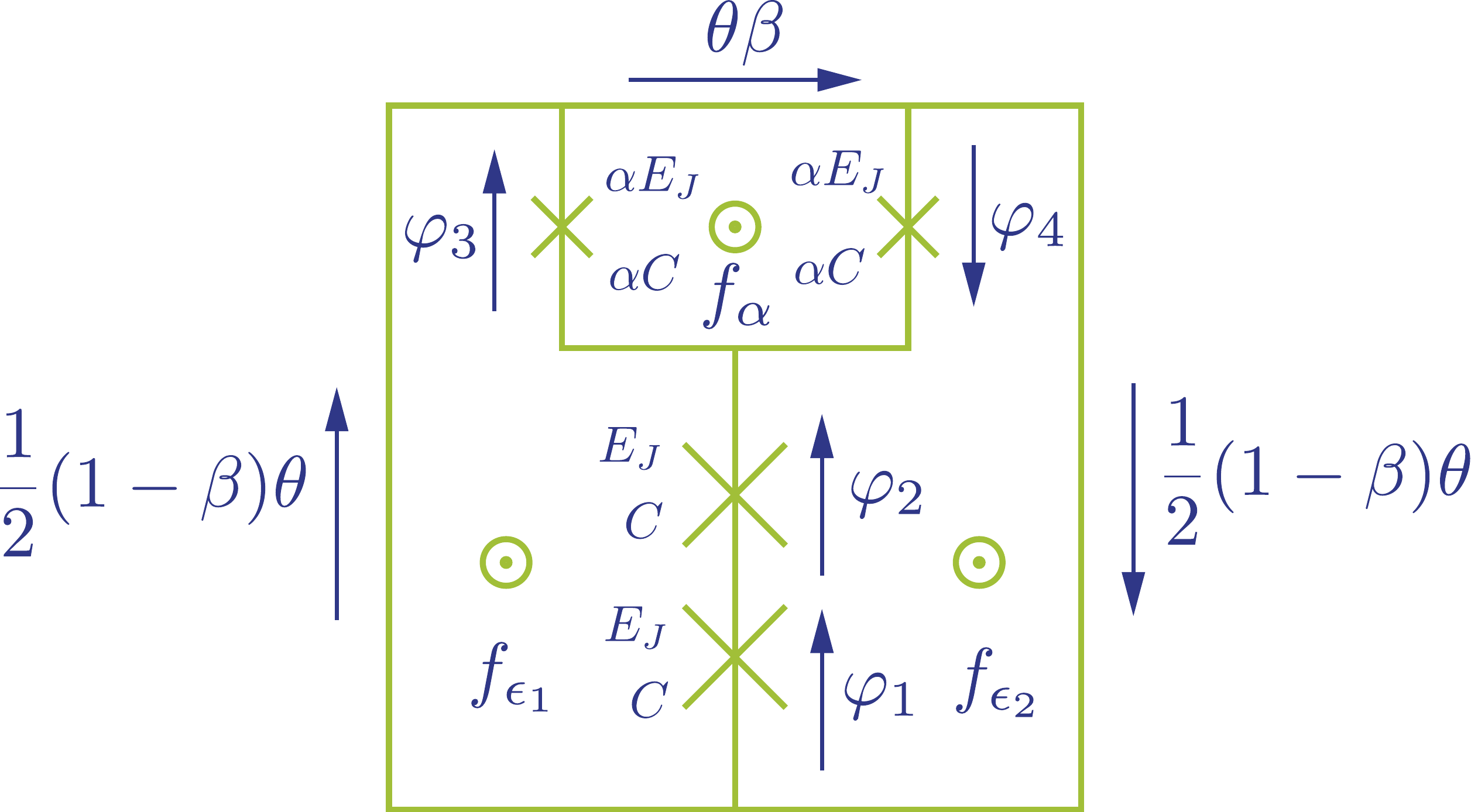}
\caption{Circuit representation of a gap tunable flux qubit. The sign $\times$ denotes the Josephson junction. The long arrows denote the current direction. The phase accumulated along the main trap-loop is denoted by $\theta$. The parameter $\beta$ denotes the ratio between the circumference of the $\alpha$-loop  and that of the main trap-loop. The $f_{\alpha}=\Phi_{\alpha}/Phi_{0}$ denotes the reduced magnetic flux threading the $\alpha$ loop, and $f_1=\Phi_{1}/\Phi_{0}$ and  $f_2=\Phi_{2}/\Phi_{0}$  denote the reduced magnetic fluxes through the left and right parts of the main loop, respectively. $E_{J}$ and $C_{J}$ denote the Josephson energy and capacitance, respectively. The phase difference across each junction is denoted by $\varphi_i$.}\label{fig:qubit}
\end{figure}

A Gap-tunable flux qubit~\cite{Paauw2009,Paauw2009Thesis,Schwarz2013}, as shown in~\figref{fig:qubit}, replaces
the $\alpha$ junction of a three-junction flux qubit~\cite{Orlando1999} with a SQUID. This introduces an external
controllable flux $f_{\alpha}$ to tune $\alpha$ in situ, thus changing the gap (qubit frequency).
To keep $f_{\alpha}$ from affecting the biasing flux $f_\epsilon$ threading the main loop, a gradiometric design
is adopted. In this $8$-shaped design, the central current of the three-junction flux qubit is split into two
opposite running currents. Because the magnetic flux generated by these two currents now cancel each other in the main loop,
independent control over both flux $f_{\alpha}$ and $f_{\epsilon}$ is ensured. Different types of tunable couplings
can also be created via different loops. For example, longitudinal coupling between two qubits ($\sigma_l^z\sigma_{l+1}^z$)
can be realized by inductively coupling two qubits through their $f_{\alpha}$-loop~\cite{Billangeon2015}. 
In this paper, we focus on the transverse coupling ($\sigma_l^x\sigma_{l+1}^x$) through the $f_{\epsilon}$-loop.

We follow the notations in Refs.~\cite{Paauw2009Thesis,Billangeon2015}, as shown in \figref{fig:qubit}.
We assume the phase accumulated along the main trap-loop is $\theta$. Here $\beta$ denotes the ratio between the circumference of the $\alpha$-loop to the main trap-loop. The magnetic frustration threading the corresponding loop
is denoted by $f_i$. The phase difference across the junction is $\varphi_i$.

Then the flux quantization conditions for the main trap-loop, $\alpha$-loop, $f_{\epsilon_{1}}$-loop,
$f_{\epsilon_{2}}$-loop are
\begin{eqnarray}
\theta+2\pi(f_{\epsilon_{1}}+f_{\epsilon_{2}}+f_{\alpha})=2\pi N\nonumber \\
\varphi_{3}+\varphi_{4}+\beta\theta+2\pi f_{\alpha}=2\pi N_{\alpha}\nonumber \\
\frac{1}{2}(1-\beta)\theta-\varphi_{3}-\varphi_{2}-\varphi_{1}+2\pi f_{\epsilon_{1}}=2\pi N_{1}\nonumber \\
\frac{1}{2}(1-\beta)\theta+\varphi_{1}+\varphi_{2}-\varphi_{4}+2\pi f_{\epsilon_{2}}=2\pi N_{2},
\end{eqnarray}
where $N_i$ is the number of trapped fluxoids.

Using above conditions,
$\varphi_{3}$, $\varphi_{4}$ can be expressed in terms of $\varphi_{1}$, $\varphi_{2}$,
\begin{eqnarray}
\varphi_{3}=-\pi[\beta(N-f_{\Sigma})+f_{\alpha}]-(\varphi_{1}+\varphi_{2})-\pi(n-f_{\epsilon})+\pi N_{\alpha},\nonumber \\
\varphi_{4}=-\pi[\beta(N-f_{\Sigma})+f_{\alpha}]+(\varphi_{1}+\varphi_{2})+\pi(n-f_{\epsilon})+\pi N_{\alpha},\nonumber \\
\end{eqnarray}
where $f_{\Sigma}=f_{\epsilon_{1}}+f_{\epsilon_{2}}+f_{\alpha}$, $f_{\epsilon}=f_{\epsilon_{1}}-f_{\epsilon_{2}}$, $N=N_{1}+N_{2}+N_{\alpha}$, $n=N_{1}-N_{2}$. For simplicity, we assume $N_{\alpha}=0$ in the following analysis.

Following the standard circuit quantization process~\cite{Devoret1995,Wendin2005},
the charging energy of the capacitor represents the kinetic energy, while the Josephson energy represents the potential energy. Then the Lagrangian of the circuit in terms of $\varphi_{1}$, $\varphi_{2}$ is
\begin{eqnarray}
\mathscr{L}&(\dot{\varphi_{i}},\varphi)&=\nonumber\\ &(\frac{\hbar}{2e})^{2}&[(1+2\alpha)\frac{C}{2}(\dot{\varphi_{1}}^{2}+\dot{\varphi_{2}}^{2})+2\alpha C\dot{\varphi_{1}}\dot{\varphi_{2}}]\nonumber \\
	& -E_{J} & (2(1+\alpha)-\cos\varphi_{1}-\cos\varphi_{2}-\nonumber \\
    &2\alpha\cos&\{\pi[\beta(N-f_{\Sigma})+f_{\alpha}]\}\cos[(\varphi_{1}+\varphi_{2})+\pi(n-f_{\epsilon})]).\nonumber
    \\
\end{eqnarray}
The canonical momentum $p_i$ conjugated to coordinate $\varphi_i$ is

\bea
p_{1}=\frac{\partial L}{\partial\dot{\varphi_{1}}}=(\frac{\hbar}{2e})^{2}[(1+2\alpha)C\dot{\varphi_{1}}+2\alpha C\dot{\varphi_{2}}],\nonumber \\
p_{2}=\frac{\partial L}{\partial\dot{\varphi_{2}}}=(\frac{\hbar}{2e})^{2}[(1+2\alpha)C\dot{\varphi_{2}}+2\alpha C\dot{\varphi_{1}}].
\eea

The Hamiltonian is related to the Lagrangian by Legendre transformation
\begin{eqnarray}
&\mathscr{H}(p_{i},\varphi_{i})&=\sum p_{i}\dot{\varphi}_{i}-\mathscr{L}\\
	&=& \frac{4E_C}{(1+4\alpha)}[(1+2\alpha)n_{1}^{2}-4\alpha n_{1}n_{2}+(1+2\alpha)n_{2}^{2}]\nonumber \\
	&+E_{J}(&2(1+\alpha)-\cos\varphi_{1}-\cos\varphi_{2}\nonumber \\
   &-2\alpha\cos&\{\pi[\beta(N-f_{\Sigma})+f_{\alpha}]\}\cos[(\varphi_{1}+\varphi_{2})+\pi(n-f_{\epsilon})]),
	\nonumber\\
\end{eqnarray}
where charging energy of the junction is defined as $E_C=e^2/2C$.
We also introduced the number operator of Cooper-pairs on the junction capacitor $n_i=p_i/\hbar$, which can also be written as $n_i=-i\partial/\partial\varphi_i$.

The energy levels of the qubit are solved numerically by the plane-wave solutions $\Psi(\varphi_{1},\varphi_{2})=\frac{1}{2\pi}\sum_{k,l=-N}^Nc_{k,l}\exp\{-i(k\varphi_{1}+l\varphi_{2})\}$. Here $k$ ($l$) is an integer, corresponding to a state that has $k$ ($l$) Cooper pairs on junction $1$ ($2$). The total charge states is set to $N=15$.
\begin{figure*}
	\centering
	\includegraphics[width=\linewidth]{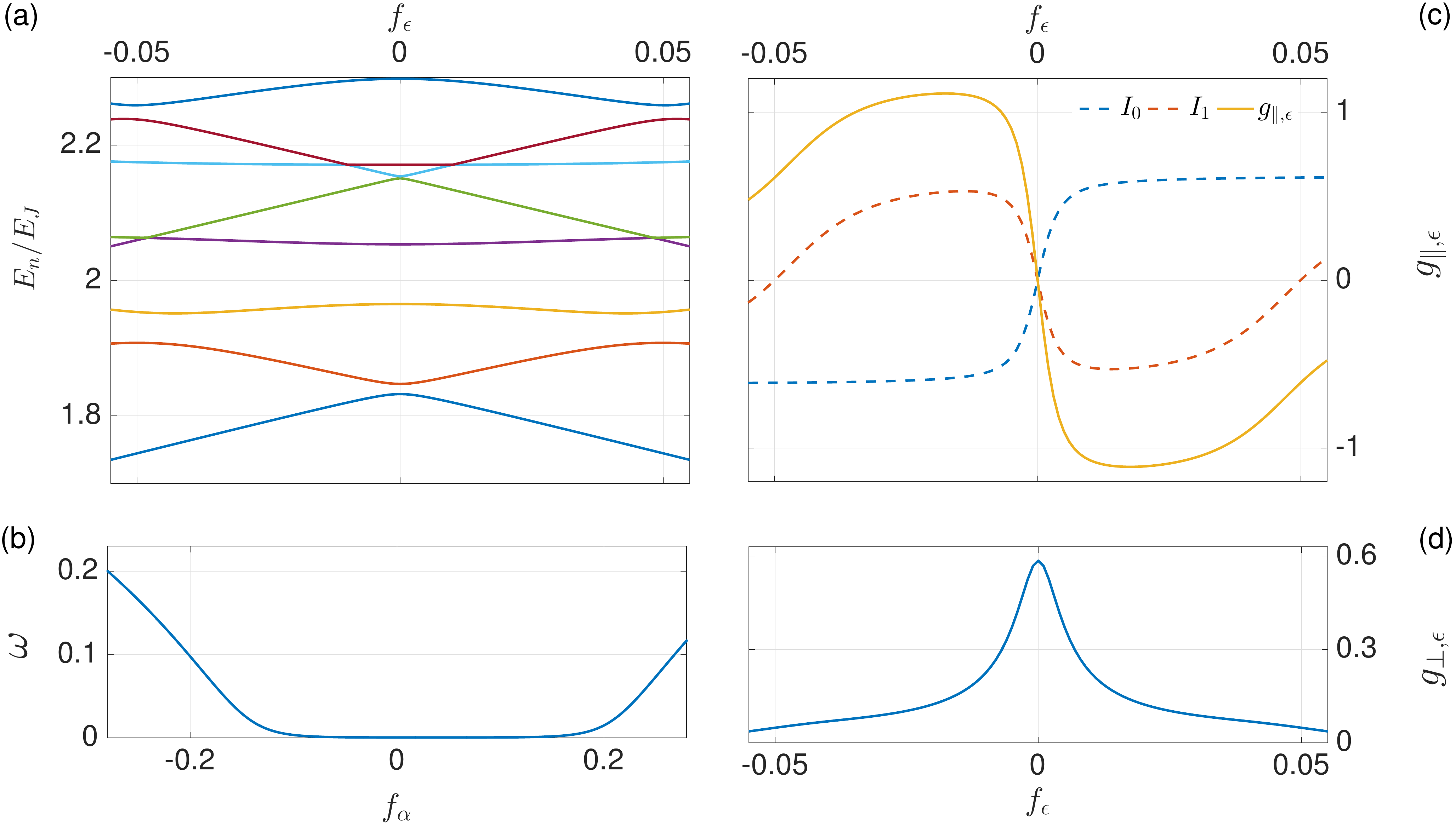}
	\caption{(a) Dependence of energy levels on the frustration $f_\epsilon$, with $f_\alpha=0.2$.
		(b) The energy gap $\omega$ as a function of $f_\alpha$ when biased at the optimal working point $f_\epsilon=0$. (c)
		Longitudinal coupling strength through $f_\epsilon$ loop, with $f_\alpha=0.2$.  (d) Transverse coupling strength through $f_\epsilon$ loop, with $f_\alpha=0.2$.
		The rest parameters are
		$E_J=1$, $E_J/E_C=50$, $\alpha=0.5$, $f_\Sigma=50f_\alpha$,
		$\beta=0.05$, $N=1$, $n=1$. 
	}\label{circuit}
\end{figure*}

The energy levels as a function of the bias flux $f_\epsilon$ are plot in~\figref{circuit} (a). At the optimal working point where $f_\epsilon=0$, the lowest two energy levels is well separated from higher excited states. The splitting of these two levels is the qubit frequency $\omega$. As shown in~\figref{circuit} (b), $\omega$ can be tuned by $f_\alpha$.

\subsection{Qubit-qubit coupling}

As shown in~\figref{fig:system}, the qubits are coupled inductively via $f_{\epsilon}$-loop. The current $I_{p,j+1}$ circulating the $f_\epsilon$ loop of the $j+1$th qubit can induce a change of flux $\delta f_{j}$ in the $f_\epsilon$ loop of qubit $j$, giving rise to the coupling term $J = MI_{pj}I_{p(j+1)}$, where the current $I_{pj}=\partial \mathscr{H}_j/\partial f_\epsilon$.  Here we assume $\delta f_{j}$ does not affect $f_{\alpha}$-loop.

As shown in the main text, the transverse coupling strength $a=g^j_{\epsilon,\bot}g^{j+1}_{\epsilon,\bot}$, with
\be
g_{\epsilon,\bot}=\langle e|\frac{\partial \mathscr{H}}{\partial f_{\epsilon}}|g\rangle.
\ee
For simplicity we have dropped the superscript $j$.

Similarly the longitudinal coupling $\lambda\sigma_j^z \sigma_{j+1}^z$ is also possible, where $\lambda=g^j_{\epsilon,\|}g^{j+1}_{\epsilon,\|}$, with
\be
g_{\epsilon,\|}=\langle +|\frac{\partial \mathscr{H}}{\partial f_{\epsilon}}|-\rangle.
\ee
Here $|\pm\rangle=(|e\rangle \pm |g\rangle)/\sqrt{2}$.

Note because of the gradiometric geometry, the flux in the main trap loop is insensitive to a homogeneous magnetic field.
Thus only the asymmetrical flux $f_\epsilon=f_{\epsilon_{1}}-f_{\epsilon_{2}}$ contributes in the coupling.

We plot $g_{\epsilon,\|}$ and $g_{\epsilon,\bot}$ as a function of the bias flux $f_\epsilon$ in~\figref{circuit} (c) and (d).
At the optimal working point $f_\epsilon=0$, $g_{\epsilon,\bot}$ is the largest, while $g_{\epsilon,\|}=0$. This is consistent with the symmetry of the flux qubit. The potential energy is symmetrical about the optimal working point,
the loop currents in the ground state $I_{0}=\langle g|\frac{\partial \mathscr{H}}{\partial f_{\epsilon}}|g\rangle$ ,  and in the first excited $I_{1}=\langle e|\frac{\partial \mathscr{H}}{\partial f_{\epsilon}}|e\rangle$
are zero. Thus when biasing at the optimal point, the qubit is first-order insensitive to dephasing noise.

To summarize, when the qubits are all biased at the optimal working point, only transverse coupling are present. Even there is a residual longitudinal coupling, this will only add fluctuations to the diagonal term of the matrix in the single-spin excitation subspace. However, the previous result shows the chain is robust under fluctuations.
To create alternating coupling pattern of the SSH model, one can vary the qubit spacing, that will change the mutual inductance $M$.

%

\end{document}